\documentstyle [11pt]{article}
\def\E{{\cal E}}

\def\K{{\cal K}}

\def\b{\begin{equation}}
\def\e{\end{equation}}
\def\bd{\begin{displaystyle}}
\def\ed{\end{displaystyle}}
\def\ba{\begin{array}}
\def\ea{\end{array}}

\def\bee{\begin{enumerate}}
\def\eee{\end{enumerate}}

\def\bes{\begin{eqnarray*}}
\def\ees{\end{eqnarray*}}
\def\be{\begin{eqnarray}}
\def\ee{\end{eqnarray}}

\begin{document}
\title{{\textbf{de Sitter field equations from quadratic curvature gravity: A group theoretical approach }}}
\author{M. Dehghani \thanks{e-mail: m.dehghani@razi.ac.ir }$\;$ and M.R. Setare \thanks{e-mail: rezakord@ipm.ir }}
 \maketitle \centerline{\it Department of
Physics, Razi University, Kermanshah, Iran.}

\begin{abstract}

In this paper, the linearized field equations related to the
quadratic curvature gravity theory have been obtained in the
four-dimensional de Sitter (dS) space-time. The massless spin-2
field equations have been written in terms of the Casimir operators
of dS group making use of the ambient space notations. By imposing
some simple constraints, arisen from group theoretical
interpretation of the field equations, a new four-dimensional
Gauss-Bonnet (GB)-like action has been introduced with the related
field equations transforming according to the unitary irreducible
representations (UIR's) of dS group. Since, the field equations
transform according to the UIR's of dS group, the GB-like action, we
just obtained, is expected to be a successful model of modified
gravity. For more clarifying, the gauge invariant field equations
have been solved in terms of a gauge-fixing parameter ${\cal{C}}$.
It has been shown that the solution can be written as the
multiplication of a symmetric rank-2 polarization tensor and a
massless minimally coupled scalar field on dS space. The
Gupta-Bleuler quantization method has been utilized and the
covariant two-point function has been calculated in terms of the
massless minimally coupled scalar two-point function, using the
ambient space notations. It has been written in terms of dS
intrinsic coordinates from the ambient space counterpart. The
two-point functions are dS invariant and free of any theoretical
problems. It means that the proposed model is a successful model of
modified gravity and it can produce significant results in the
contexts of classical theory of gravity and quantum gravity toy
models.
\end{abstract}

{\it Proposed PACS numbers}: 04.62.+v, 03.70+k, 11.10.Cd, 98.80.H,
98.80.Jk

\newpage

\setcounter{equation}{0}
\section{Introduction}

Among the reasons it should be interesting to study the physics of
dS and asymptotically dS spacetimes we may mention that: (1) Recent
analysis of astronomical data indicates that there is a positive
cosmological constant and we live in a universe which will look like
dS spacetime in the future \cite{ds1, ds2, ds3}. (2) The interesting
proposal of defining, in a manner analogous to the AdS/CFT
correspondence, the dS/CFT correspondence has been suggested
recently. According to dS/CFT correspondence, there is a dual
between quantum gravity on dS space and a Euclidean conformal field
theory on the boundary of dS space \cite{ds4}. Therefore,
understanding of quantum field theory on dS space-time is of
considerable interest.

Although the Einstein tensorial theory of gravitation, known today
as general relativity, was more successful to pass observational
tests and became the standard theory of gravitation, according to
the recent cosmological observations it seems that this theory may
be incomplete. The most important failures was the inability to
describe accelerating expansion of the universe \cite{prl1, prl2,
prl3}, cosmic microwave background anisotropies \cite{sp1, sp2},
large scale structure formation \cite{te1, te2}, baryon oscillations
\cite{ei} and weak lensing \cite{ja}, etc. One of the main
approaches for explanation of these phenomena is to modify Einstein
general relativity. In this regards, various modifications of
Einstein gravity were proposed in the literatures. Among them
Lovelock gravity \cite{lov1, lov2, lov3}, braneworld scenario
\cite{bra1, bra2, bra3}, scalar-tensor theories \cite{jor1, jor2,
jor3}, f(R) gravity \cite{far1}-\cite{far4} and, in particular,
non-linear gravity theories or higher-order theories of gravity have
provided interesting results \cite{cap}-\cite{ no4}. Some of the
models are based on gravitational actions which are non-linear in
the Ricci scalar, the Riemann and Ricci invariants, known as
quadratic curvature gravity theory, or contain terms involving
combinations of derivatives of the Ricci scalar
\cite{ke1}-\cite{mag}.

Nowadays, the quadratic curvature gravity theories and their
applications have been the subject of many interesting works and a
lot of papers have been appeared in which the usual theory of
gravity is extended quadratically making use of the curvature
invariants (for example see \cite{qe1}-\cite{fan}).
Quadratic curvature gravity is a natural generalization on
Einstein's gravity. That includes higher order derivatives of
metric. But due to the presence of these higher order derivatives,
massive ghosts appear. However, higher order models of modified
gravity play an important role in high energy physics. These certain
classes of higher order gravity theories are known as the quantum
gravity toy models \cite{a, b}.

We already know that the gravitational field equations in the linear
approximation describe a massless spin-2 particle (the graviton, if
it exists) which propagates on the background space-time. Following
the Wigner's theorem, a linear gravitational field should transform
according to the UIR's of the symmetric group of the background
space-time. In the previous paper \cite{de22}, making use of an
action containing quadratic form of both the Ricci scalar and the
Ricci tensor, we obtained the \textbf{physical} massless spin-2
field equations and showed that it can be associated with UIR's of
dS group with a suitable choice of constant coefficients. Also we
obtained the massless rank-2 tensor field and the related two-point
function as the solutions to the \textbf{physical} field equations.

The main purpose of this paper is to introduce a new extension of
the Hilbert-Einstein action of the gravitational field. This action
is composed by taking a linear combination of the squared Ricci
scalar, the Ricci and Riemann curvature invariants. To this end, we
consider the linearized form of the corresponding field equation
around dS background. Recasting these linearized field equations in
terms of Casimir operators dS group one is led to a new GB-like
quadratic action in the four dimensions. Since the field equations,
stem from the new action we just obtained, are correspond to the
UIR's of dS group it is a successful model of modified gravity and
can produce reasonable results. For more clarifying, we
solve the field equations in terms of the massless minimally coupled
scalar field in dS space. Also we obtain the related two-point
function and show that it can be written in terms of the massless
minimally coupled scalar two-pint function on dS space. The results
of the calculations are free of any theoretical problems and confirm
the validity and successfulness of the proposed model of modified
gravity.

This paper is structured as follows. In section-2, the most general
form of the quadratically-extended gravitational action, as the
generalization of the Hilbert-Einstein action, has been introduced
and corresponding linearized field equations have been obtained in
terms of the intrinsic dS coordinates as the background. In
section-3, after a brief review of the ambient space notations and
Casimir operators of dS group, the full field equations have been
written in terms of Casimir operators of dS group. By imposing some
simple conditions a new GB-like action has been obtained with the
field equations which are correspond to the UIR's of dS group. In
order to illustrate the validity and successfulness of the action,
the related field equations have been solved in terms of the
gauge-fixing parameter ${\cal{C}}$ making use of the ambient space
notations. The solution is written as the multiplication of a
symmetric generalized polarization rank-2 tensor and a massless
minimally coupled scalar field on dS space. Also, making use of the
Gupta-Bleuler quantization method, we calculated the corresponding
two-point function in terms of the massless minimally coupled scalar
two-point function using the ambient space formalism. The
dS-invariant graviton two-point function has been written in terms
of 4-dimensional intrinsic coordinates from its ambient space
counterpart. The results are summarized and discussed in section-4.
Some useful mathematical relations and details of derivations have
been given in appendices.

\setcounter{equation}{0}
\section{Linear dS field equations}

The gravitational action for the quadratic curvature gravity theory
in the 4-dimensional dS space-time with the positive cosmological
constant $\Lambda=3H^2$ can be written as \b I=\frac{1}{16\pi G}\int
d^4 x\sqrt{-g}\left[a_0({\cal{R}}-2\Lambda )
+a{\cal{R}}^2+b{\cal{R}}^{ab}{\cal{R}}_{ab}+c{\cal{R}}^{abcd}{\cal{R}}_{abcd}
\right],\e where ${\cal{R}}_{abcd}$ is the Riemann tensor,
${\cal{R}}_{ab}$ is the Ricci tensor and
${\cal{R}}=g^{ab}{\cal{R}}_{ab}$ is the Ricci scalar of the
space-time under consideration. $a_0$, $a$, $b$ and $c$ are constant
coefficients. The coefficients $a$, $b$ and $c$ are positive having
dimension of $(\mbox{Length})^2$. The theory described by this
action is referred to as fourth-order gravity since it leads to
fourth order equations. Numerous papers have been devoted to the
study of fourth-order gravity theories. Note that the Gauss-Bonnet
action \b\frac{1}{16\pi G}\int d^4 x\sqrt{-g}\left(
{\cal{R}}^2-4{\cal{R}}^{ab}{\cal{R}}_{ab}+{\cal{R}}^{abcd}{\cal{R}}_{abcd}\right),\e
is a total divergence. It does not contribute to the field equations
of 4-dimensional space-times.

By varying the action (2.1) with respect to the metric tensor
$g_{ab}$ the modified gravitational field equations can be obtained
as \cite{fan} \b E_{ab}= a_0 \E_{ab}^{(0)}+a \E_{ab}^{(1)}+b
\E_{ab}^{(2)}+c\E_{ab}^{(3)}=0,\e where
$\E_{ab}^{(0)}=G_{ab}+\Lambda g_{ab}$ and $
G_{ab}={\cal{R}}_{ab}-\frac{1}{2}{\cal{R}}g_{ab}$ is the Einstein
tensor and   \b \E_{ab}^{(1)}=2{\cal{R}}{\cal{R}}_{ab}-2\nabla_a
\nabla_b {\cal{R}}-\frac{1}{2}g_{ab}({\cal{R}}^2-4\Box {\cal{R}}),\e
\b \E_{ab}^{(2)}=\Box
{\cal{R}}_{ab}-\nabla_a\nabla_b{\cal{R}}+2{\cal{R}}_{acbd}{\cal{R}}^{cd}
-\frac{1}{2}g_{ab}({\cal{R}}^{cd}{\cal{R}}_{cd}-\Box{\cal{R}}),\e
\b\E_{ab}^{(3)}=4\Box{\cal{R}}_{ab}-2\nabla_a\nabla_b{\cal{R}}-4
{\cal{R}}_{ac}
 {\cal{R}}^c_b+4{\cal{R}}_{acbd}
{\cal{R}}^{cd}+2{\cal{R}}_{acde}{\cal{R}}_b^{\;
cde}-\frac{1}{2}g_{ab}{\cal{R}}_{cdef}{\cal{R}}^{cdef},\e and
$\Box\equiv\nabla^a\nabla_a=g^{ab}\nabla_a\nabla_b$ is the
d'Alembertian operator.

We use the background field method to obtain the linearized form of
the field equations (2.3). According to the background field method,
originally developed by Christian Fronsdal \cite{fr}, one can assume
$g_{ab}=g_{ab}^{(BG)}+h_{ab},$ in which $g_{ab}^{(BG)}$ is the
background metric and $h_{ab}$ are its small fluctuations. Indices
are raised and lowered by the background metric. We suppose that
$g_{ab}^{(BG)}=g_{ab}^{(ds)}\equiv \tilde{g}_{ab}$. So we have \b
g_{ab}\simeq\tilde{g}_{ab}+h_{ab}\;\;\;\;\;\;\;
\mbox{and}\;\;\;\;\;\;\;g^{ab}\simeq\tilde{g}^{ab}-h^{ab}.\e

The metric $\tilde{g}_{ab}$ is a solution to Einstein's field
equations with the positive cosmological constant $\Lambda=3H^2$: \b
\tilde{R}_{ab}-\frac{1}{2}\tilde{R}\tilde{g}_{ab}+3H^2
\tilde{g}_{ab}=0. \e

Making use of the approximations given in Eq.(2.7), in Eq.(2.3), we
have \b
(\E_{ab}^{(0)})_L=\frac{1}{2}(\nabla_{a}\nabla^{c}h_{bc}+\nabla_{b}\nabla^{c}h_{ac}-\Box
 h_{ab}-\nabla_{a}\nabla_{b}h'+2H^2h_{ab}) +\frac{1}{2}\tilde{g}_{ab}(\Box
h'-\nabla_{c}\nabla_{d}h^{cd}+H^2h'),\e in which $h'=h_{a}^{a}$ is
the trace of $h_{ab}$ with respect to the background metric and
$\nabla^b$ is the background covariant derivative. Making use of the
relations given in appendix-A it is easy to obtain the linearized
field equations in dS space as
$$(\E_{ab}^{(1)})_L=+12H^2\left(\nabla_{a}\nabla^{c}h_{bc}+\nabla_{b}\nabla^{c}h_{ac}-\Box
h_{ab}\right)-2\nabla_a\nabla_b\left(\nabla_c\nabla_d h^{cd}-\Box
h'+3H^2h'\right)$$
\b+24H^4h_{ab}-2\tilde{g}_{ab}\left(3H^2\nabla_{c}\nabla_{d}h^{cd}+3H^4h'-\Box
\nabla_{c}\nabla_{d}h^{cd}+\Box^2h'\right),\e
$$(\E_{ab}^{(2)})_L=\frac{1}{2}\left[\Box \left( \nabla_a \nabla_ch^c_b+\nabla_b \nabla_ch^c_a \right)
-2H^2\Box h_{ab}-\Box^2h_{ab}+ \nabla_a \nabla_b\Box h'\right]$$
$$+2H^2\left( \nabla_a \nabla_ch^c_b+\nabla_b \nabla_ch^c_a\right) -\nabla_a \nabla_b \nabla_{c}\nabla_{d}h^{cd}
-3H^2 \nabla_a \nabla_b h'+4H^4 h_{ab}$$
\b+\frac{1}{2}\tilde{g}_{ab}\left(2H^2\nabla_{c}\nabla_{d}h^{cd}-2H^4h'+7H^2\Box
h'+\Box \nabla_{c}\nabla_{d}h^{cd} -\Box^2h'\right),\e
$$(\E_{ab}^{(3)})_L=2\Box \left( \nabla_a \nabla_ch^c_b+\nabla_b
\nabla_ch^c_a \right) +8H^2\Box h_{ab}-2\Box^2h_{ab}-8H^4h_{ab}-6H^2
\nabla_a \nabla_b h'$$ \b-4H^2\left( \nabla_a \nabla_ch^c_b+\nabla_b
\nabla_ch^c_a\right) -2\nabla_a \nabla_b \nabla_{c}\nabla_{d}h^{cd}
+2H^2\tilde{g}_{ab}\left(\nabla_{c}\nabla_{d}h^{cd}-\Box h'+H^2
h'\right) .\e Now, making use of Eqs.(2.9)-(2.12) and Eq.(2.3) we
have \b (E_{ab})_L=
a_0(\E_{ab}^{(0)})_L+a(\E_{ab}^{(1)})_L+b(\E_{ab}^{(2)})_L+c(\E_{ab}^{(3)})_L=0.\e
Eq.(2.13) is the linearized quadratically-extended gravitational
field equations in dS background, which has been written in terms of
the intrinsic coordinates $X_a$ of the 4-dimensional dS space-time.
The linear field equations (2.13) can be rewritten in the following
explicit form
$$(E_{ab})_L=-\left(\frac{a_0}{2}+12aH^2+bH^2-8cH^2\right)\Box
h_{ab}+H^2(a_0+24aH^2+4bH^2-8cH^2)h_{ab}$$
$$-\left(\frac{b}{2}+2c\right)\Box^2
h_{ab}+\left( 2a+\frac{b}{2}\right)\nabla_a\nabla_b \Box h'-\left(
\frac{a_0}{2}+6aH^2+3bH^2+6cH^2\right)\nabla_a\nabla_b h' $$
$$+\left(\frac{a_0}{2}+12aH^2+2bH^2+\frac{b}{2}\Box-4cH^2+2c\Box\right)\left( \nabla_{a}\nabla^{c}h_{bc}+\nabla_{b}\nabla^{c}h_{ac} \right)
$$
$$+\frac{1}{2}\tilde{g}_{ab}\left[\left(-a_0-12aH^2-2bH^2+4cH^2\right)\nabla_{c}\nabla_{d}h^{cd}+\left(a_0-bH^2-4cH^2\right)\Box
h'\right.$$
 $$ \left. +\left(a_0-12aH^2-2bH^2+4cH^2\right)H^2h'+\left(4a+b\right)
\left(\Box\nabla_{c}\nabla_{d}h^{cd}-\Box^2 h'\right)\right]$$
\b-(2a+b+2c)\nabla_a\nabla_b \nabla_{c}\nabla_{d}h^{cd} =0.\e It is
easy to show that the massless spin-2 dS field equations (2.14) is
invariant under the following gauge transformations $$
h_{ab}\rightarrow h^{(gt)}_{ab}=h_{ab}+\nabla_a \Lambda _b+\nabla_b
\Lambda _a, $$ where $\Lambda _a$ is an arbitrary four-vector.

Letting $a=c=\alpha_g$ and $b=-4\alpha_g$, where $\alpha_g$ is a
constant coefficient with dimension of $(\mbox{Length})^2$, the
proposed gravitational theory (2.1) reduces to the
Einstein-Gauss-Bonnet theory as \b\frac{1}{16\pi G}\int d^4
x\sqrt{-g}\left[a_0({\cal{R}}-2\Lambda )+\alpha_g\left(
{\cal{R}}^2-4{\cal{R}}^{ab}{\cal{R}}_{ab}+{\cal{R}}^{abcd}{\cal{R}}_{abcd}\right)\right],\e
and the linearized field equations (2.14) reduces to $(E_{ab})_L=a_0
(\E_{ab}^{(0)})_L$. It means that the Gauss-Bonnet action is a total
divergence and does not contribute to the field equations even if in
its linear approximation. It is evident that the theory with these
constraints is not of interest here.

 The Minkowskian correspondence of the theory can be obtained by
taking the zero curvature limit (i.e. $H \rightarrow 0$) of
Eq.(2.14); it is,
$$2(E_{ab})_L^{Mink.}=-\left(b+4c\right)\Box^2 h_{ab}-a_0\Box
h_{ab}+\left( 4a+b\right)\partial_a\partial_b \Box h'-
a_0\partial_a\partial_b h' $$
$$+\left(a_0+b\Box+4c\Box\right)\left( \partial_{a}\partial^{c}h_{bc}+\partial_{b}\partial^{c}h_{ac} \right)
-2(2a+b+2c)\partial_a\partial_b \partial_{c}\partial_{d}h^{cd}$$
 \b\eta_{ab}\left[a_0\Box h'-a_0\partial_{c}\partial_{d}h^{cd}+(4a+b)
\left(\Box\partial_{c}\partial_{d}h^{cd}-\Box^2
h'\right)\right]=0.\e where $\eta_{ab}$ is the metric and
$\Box=\eta_{ab}\partial^a\partial^b=\partial^a\partial_a$ is the
wave operator in the flat space. The linearized Minkowskian field
equations (2.16) is invariant under the following gauge
transformations $$ h_{ab}\rightarrow h^{(gt)}_{ab}=h_{ab}+\partial_a
\lambda _b+\partial_b \lambda _a,$$ for the arbitrary four-vector
$\lambda _a$.

Now, we calculate the trace of the field equations (2.14). In terms
of the dimensionless constants $A=aH^2,\; B=bH^2\;\mbox{and}\;
C=cH^2$, that is
$$ (E')_L=\tilde{g}^{ab}(E_{ab})_L=\left(a_0-18A-6B-6C \right)\Box h'+3H^2a_0 h'-2\left(3A+B+C
\right)H^{-2}\Box^2h'$$ \b-a_0\nabla_c\nabla_d h^{cd}+2
\left(3A+B+C\right)H^{-2}\Box \nabla_c\nabla_d h^{cd}=0,\e which
immediately reads as \b a_0=0, \;\;\;\;\;\;\;\;\;\;\;\;\; \mbox{and}
   \;\;\;\;\;\;\;\;\;\;\;\;\;   3A+B+C=0.\e
Making use of these relations in the field equations (2.14) we
obtain
$$-2(E_{ab})_L=\left( B+4C\right)\left[H^{-2}\Box^2h_{ab}-H^{-2}\Box \left(
\nabla_{a}\nabla^{c}h_{bc}+\nabla_{b}\nabla^{c}h_{ac} \right)-6\Box
h_{ab}+8H^2h_{ab}\right.$$
$$+4\left(
\nabla_{a}\nabla^{c}h_{bc}+\nabla_{b}\nabla^{c}h_{ac}
\right)+2\nabla_a\nabla_b h'+\frac{2}{3}H^{-2}\nabla_a\nabla_b
\nabla_{c}\nabla_{d}h^{cd}+\frac{1}{3}H^{-2}\nabla_a\nabla_b \Box
h'$$ \b \left.-\tilde{g}_{ab}\left(2\nabla_{c}\nabla_{d}h^{cd}-\Box
h'+2 H^2h'-\frac{1}{3}H^{-2}
\left(\Box\nabla_{c}\nabla_{d}h^{cd}-\Box^2
h'\right)\right)\right]=0.\e Since $B\neq-4C$, Eq.(2.19) can be
rewritten as

$$H^{-2}\Box^2h_{ab}-H^{-2}\Box \left(
\nabla_{a}\nabla^{c}h_{bc}+\nabla_{b}\nabla^{c}h_{ac} \right)-6\Box
h_{ab}+8H^2h_{ab}$$
$$+4\left(
\nabla_{a}\nabla^{c}h_{bc}+\nabla_{b}\nabla^{c}h_{ac}
\right)+2\nabla_a\nabla_b h'+\frac{2}{3}H^{-2}\nabla_a\nabla_b
\nabla_{c}\nabla_{d}h^{cd}+\frac{1}{3}H^{-2}\nabla_a\nabla_b \Box
h'$$ \b -\tilde{g}_{ab}\left[2\nabla_{c}\nabla_{d}h^{cd}-\Box h'+2
H^2h'-\frac{1}{3}H^{-2} \left(\Box\nabla_{c}\nabla_{d}h^{cd}-\Box^2
h'\right)\right]=0.\e

 \setcounter{equation}{0}
\section{The field equations in the ambient space}
In this section we try to rewrite the linearized field equations
(2.20) making use of the five-dimensional ambient space formalism.
For this purpose, at firs we review briefly the ambient space
notations and remind the Casimir operators of dS group. We consider
a symmetric and transverse (i.e. $x\cdot\K(x)=0$) tensor field
$\K_{\alpha\beta}(x)$ in ambient space notations. It is related to
the ``intrinsic'' field $h_{ab}(X)$ through the following tensorial
transformation rule \cite{derota, di, gagata} \b
h_{ab}(X)=\frac{\partial x^{\alpha}}{\partial X^{a}}\frac{\partial
x^{\beta}}{\partial X^{b}}\K_{\alpha\beta}(x(X)). \e The covariant
derivative in the ambient space notations is defined as
\b\label{eq:cov} D_\beta T_{\alpha_1 ...\alpha_i ...\alpha_n}= \bar
\partial_\beta T_{\alpha_1 ...\alpha_i
...\alpha_n}-H^2\sum_{i=1}^nx_{\alpha_i}T_{\alpha_1 ...\beta
...\alpha_n},\e where $\bar
\partial$ is  tangential (or transverse) derivative in
dS space \b \bar
\partial_\alpha=\theta_{\alpha \beta}\partial^\beta=
\partial_\alpha  +H^2x_\alpha x\cdot\partial,\;\;\;x\cdot\bar \partial=0,\e
and $\theta_{\alpha \beta}$ can be written in terms of the ambient
space metric $\eta_{\alpha \beta}=$diag$(1,-1,-1,-1,-1)$ as
$\theta_{\alpha \beta}=\eta_{\alpha \beta}+H^2x_{\alpha}x_{ \beta}$.
 It is easily shown that the dS metric $\tilde{g}_{ab}$ corresponds to
the transverse projector $\theta_{\alpha\beta}$, that is, \b
\tilde{g}_{ab}(X)=\frac{\partial x^{\alpha}}{\partial
X^{a}}\frac{\partial x^{\beta}}{\partial
X^{b}}\theta_{\alpha\beta}(x). \e The two Casimir operators of dS
group are labeled by $Q^{(1)}_s$ and $Q^{(2)}_s$ where the subscript
$s$ reminds the rank of tensors in consideration. The action of the
Casimir operators $Q_1^{(1)}$ and $Q_2^{(1)}$  on $K$ and ${\cal K}$
respectively can be written in the more explicit form
\begin{equation}\label{eq:act}
Q_1^{(1)} K(x)=\left(Q_{0}^{(1)}-2\right) K(x)+2x
\bar{\partial}\cdot K(x)+2H^2 x\;x\cdot K(x)-2  \bar{\partial}\;
x\cdot K(x),\end{equation}
\begin{equation}\label{eq:act}
 Q_2^{(1)}{\cal
K}(x)=\left(Q_{0}^{(1)}-6\right){\cal K}(x)+2\eta {\cal K}'+2{\cal
S} x\bar{\partial}\cdot{\cal K}(x)-2{\cal S}  \bar{\partial}
x\cdot{\cal K}(x)+2H^2{\cal S}xx\cdot{\cal K}(x),
\end{equation} where, $Q_{0}^{(1)}=-H^{-2}(\bar\partial)^2$
is the scalar Casimir operator. The symmetrizer ${\cal S}$ is
defined for two vectors $\xi_{\alpha}$ and $\omega_{\beta}$ by
${\cal
S}(\xi_{\alpha}\omega_{\beta})=\xi_{\alpha}\omega_{\beta}+\xi_{\beta}\omega_{\alpha}$.
$\K'=\eta^{\alpha\beta}\K_{\alpha\beta}$ is the trace of the tensor
$\K$. The readers are referred to \cite{derota} and references
therein fore more details.

Making use of the ambient space formalism, the field equations
(2.20) can be written as (appendix-B)
\b(Q_2^{(1)}+4)(Q_2^{(1)}+6)\K+(Q_2^{(1)}+4)D_2(\partial_2.\K)=0,\e
where $D_2$ is the generalized gradient operator defined by
$D_2K={\cal{ S}}(D_1-x)K$ with
$D_{1\alpha}=H^{-2}\bar{\partial}_\alpha$ and
$\partial_2.\K=\partial.\K-H^2x\K'-\frac{1}{2}\partial\K'\;
\cite{derota}.$

Noting the following identities \cite{gagata} \b
Q_2^{(1)}D_2{\cal{A}}=D_2Q_1^{(1)}{\cal{A}} ,\e \b
\partial_2.D_2{\cal{A}}=-(Q_1^{(1)}+6){\cal{A}},\e
one can show that Eq.(3.7) is invariant under the following gauge
transformations \b \K \rightarrow \K^{(gt)}=\K+ D_2{\cal{A}},\e for
an arbitrary five-vector ${\cal{A}}$ in the ambient space. It means
that one can rewrite the field equations (3.7) in the following form
\b(Q_2^{(1)}+4)(Q_2^{(1)}+6)\K+{\cal{C}}(Q_2^{(1)}+4)D_2(\partial_2.\K)=0,\e
where ${\cal{C}}$ is a gauge-fixing parameter. It is notable that if
one let ${\cal{C}}=0$ or if one impose the physical conditions
$\bar{\partial}.\K=0=\K'$ in Eq.(3.11), the physical sector of the
theory obtains as \b(Q_2^{(1)}+4)(Q_2^{(1)}+6)\K=0,\e which
transforms according to two of the UIR's of dS group denoted by
$\Pi^{\pm}_{2,2}$ and $\Pi^{\pm}_{2,1}$ in discrete series. The
physical graviton field equations (3.12) has been solved and the
corresponding two-point function has been obtained in \cite{de22}.

Now, we are in a position to introduce a new four-dimensional
GB-like action in the form   \b I=\frac{1}{16\pi G}\int d^4
x\sqrt{-g}\left(a{\cal{R}}^2+b{\cal{R}}^{ab}{\cal{R}}_{ab}+c{\cal{R}}^{abcd}{\cal{R}}_{abcd}\right),\e
with the constraints $3a+b+c=0$ and $b\neq-4c.$ Since the
gravitational field equations related to the GB-like action (3.13)
transform according to the UIR's od dS group, we claim that it can
be a successful model of modified gravity and can produce realized
and significant physical results in the context of classical theory
of gravity such as classical black holes. For more clarifying, we
would like to solve the field equations and calculate the related
two-point functions in the following subsections.

It must be noted that if we set $c=0, \;a=-\frac{1}{3}$ and $b=1$,
in the proposed gravitational theory given by action (3.13), it
reduces to the well-known Weyl theory of gravity. As an immediate
consequence of the calculations presented here the field equations
related to the Weyl theory of gravity, in the physical state,
transform according to two of the UIR's of dS group labeled by
$\Pi^{\pm}_{2,2}$ and $\Pi^{\pm}_{2,1}$ in discrete series. This
result is not compatible with that of ref.\cite{ta1}.

\subsection{Solution to dS massless spin-2 field equations}
In this section we would like to solve the field equations (3.11) in
the ambient space notations. The general solution of to the field
equations can be constructed out by the suitable combination of a
scalar field and two vector fields. Let us introduce a rank-2 tensor
field $\K$ in terms of a five-dimensional constant vector
$Z_1=(Z_{1\alpha})$ and a scalar field $\phi_1$ and two vector
fields $K$ and $K_g$ by putting \cite{de22, derota},
\cite{gagata}-\cite{tap} \b \K=\theta\phi_1+ {\cal {S}}\bar
{Z}_1K+D_2K_g,\e where $\bar Z_{1\alpha}=\theta_{\alpha\beta}
Z^{\beta}_1$. Making use of the ansatz ($3.14$) in the field
equations and noting relations (3.8) and (3.9) with the help of the
following identity \cite{gagata} \b Q^{(1)}_2{\cal S}\bar{Z}K={\cal
S}\bar{Z}(Q^{(1)}_1+4)K-2H^2D_2(x.Z_1)K+4\theta Z_1.K ,\e we arrived
at the following equation $$\theta
\left[(Q^{(1)}_0+4)(Q^{(1)}_0+6)\phi_1+8(Q^{(1)}_0+2)Z_1.K\right]+{\cal
S}\bar{Z}_1\left[Q^{(1)}_1(Q^{(1)}_1+2)K\right]$$ $$
+D_2\left[(1-{\cal{C}})(Q^{(1)}_1+4)(Q^{(1)}_1+6)K_g-4H^2\left((Q^{(1)}_1+5)(x.Z_1K)+Z_1.D_1K-xZ_1.K\right)
\right.$$ \b \left.
-{\cal{C}}H^2(Q^{(1)}_1+4)\left(D_1\phi_1+D_1Z_1.K-H^{-2}Z_1.\bar{\partial}K+x(Z_1.K)-H^{-2}Z_1\bar{\partial}.K-5(x.Z_1)K
\right)\right]=0,\e which immediately reads as the following system
of coupled differential equations \b
Q^{(1)}_1\left(Q^{(1)}_1+2\right)K=0,\;\;\;\;\mbox{or}\;\;\;\;Q^{(1)}_1Q^{(1)}_0K=0,\;\;\;\partial
.K=0=x.K,\e \b
(Q^{(1)}_0+4)(Q^{(1)}_0+6)\phi_1+8(Q^{(1)}_0+2)Z_1.K=0,\e
$$(Q^{(1)}_1+4)(Q^{(1)}_1+6)K_g^{({\cal{C}})} =\frac{4H^2}{1-{\cal{C}}}\left[ (Q^{(1)}_1+5)(x.Z_1)K+(Z_1.D_1)K
-x(Z_1.K)\right]\;\;\;\;\;\;\;\;\;\;\;\;\;\;\;\;\;\;\;\;\;\;\;\; $$
\b+\frac{{\cal{C}}H^2}{1-{\cal{C}}}(Q^{(1)}_1+4)\left[D_1\phi_1+D_1(Z_1.K)-(Z_1.D_1)K+x(Z_1.K)-5(x.Z_1)K\right],
\;\;\;\;\; {\cal{C}}\neq 1. \e The solution to the first two
equations can be written as \cite{de22}  \b
\phi_1=-\frac{2}{3}Z_1.K, \;\;\;\;\;\;
K=\frac{\sigma}{2}\left[(\sigma+2)\bar
Z_{2}+(\sigma^2+2\sigma-2)\frac{x.Z_2}{x.\xi}\bar\xi
\right]\phi_s,\e where $\phi_s$ is the massless minimally coupled
scalar field in dS space such that \b
Q^{(1)}_0\phi_s=0,\;\;\;\;\;\phi_s=(H x.\xi)^\sigma, \;\;\;\;
\xi^2=0, \;\;\;\;\sigma=0,\;-3.\e Regarding Eq.(3.20), Eq.(3.19) can
be rewritten as
$$K_g^{({\cal{C}})} =\frac{4H^2}{1-{\cal{C}}}(Q^{(1)}_1+6)^{-1}(Q^{(1)}_1+4)^{-1}\left[ (Q^{(1)}_1+5)(x.Z_1)K+(Z_1.D_1)K
-x(Z_1.K)\right]$$
\b+\frac{{\cal{C}}H^2}{1-{\cal{C}}}(Q^{(1)}_1+6)^{-1}\left[\frac{1}{3}D_1(Z_1.K)-(Z_1.D_1)K+x(Z_1.K)-5(x.Z_1)K
\right]. \e Making use of the identities given in ref.\cite{de22},
we can write the gauge dependent vector $K_g^{({\cal{C}})}$ as $$
K_g^{({\cal{C}})}
=\frac{H^2}{3(1-{\cal{C}})}\left[(x.Z_1)K+\frac{1}{9}
D_1(Z_1.K)\right]$$
\b+\frac{{\cal{C}}H^2}{1-{\cal{C}}}(Q^{(1)}_1+6)^{-1}\left[\frac{1}{3}D_1(Z_1.K)-(Z_1.D_1)K+x(Z_1.K)-5(x.Z_1)K
\right]. \e Now, using the following identities \b
(Q^{(1)}_1+6)\left[\frac{1}{9}D_1(Z_1.K)+(x.Z_1)K\right]=6(x.Z_1)K,
\e\b(Q^{(1)}_1+6)Z.\bar\partial K=6Z.\bar\partial K+2H^2D_1Z.K,\e
\b(Q^{(1)}_1+6)D_1Z.K=6D_1Z.K,\e \b(Q^{(1)}_1+6)xZ.K=6x Z.K,\e in
Eq.(3.23) we have \b K_g^{({\cal{C}})}
=\frac{{\cal{C}}H^2}{6(1-{\cal{C}})}\left[\frac{2+{\cal{C}}}{9{\cal{C}}}
D _1(Z_1.K) + x (Z_1.K) - (Z_1.D_1) K+
\frac{2-5{\cal{C}}}{{\cal{C}}} (x. Z_1) K \right].\e
 Note that $ K_g^{({\cal{C}})} $ satisfies the conditions \b x.K_g^{({\cal{C}})}=0
,\;\;\;\;\mbox{and} \;\;\;\;\bar{\partial}.K_g^{({\cal{C}})}
=\frac{1}{3}H^2Z_1.K.\e  In addition, if one set ${\cal{C}}=0$ in
Eq.(3.28), $ K_g^{({\cal{C}})} $ reduces to its physical
correspondence given in \cite{de22}. It is obvious from Eq.(3.28)
that the simplest gage-fixed value is not ${\cal{C}}=0$. But it is
${\cal{C}}=\frac{2}{5}$, which obeys the relation
${\cal{C}}=\frac{2}{2s+1}$ and $s$ denotes spin of the particle
under consideration (graviton, if it exists). The case correspond to
${\cal{C}}=\frac{2}{2s+1}$ in named conventionally as the "Minimal
case" (see \cite{ms1} for the massless spin-1 particles and
\cite{de2} for massless spin-2 particles in the Einstein's gravity).
In the minimal case we have  \b K_g^{(\frac{2}{5})}
=\frac{2H^2}{27}\left[2 D _1(Z_1.K) + x (Z_1.K) - (Z_1.D_1) K
\right].\e

Now, Eqs.$(3.20)$ and $(3.28)$ show that one can construct the
tensor field $\K$ in terms of a ``massless'' minimally coupled
scalar field on dS space. After some calculations we can show that
\b \K^{({\cal{C}})}_{\alpha \beta}(x)= {\cal
E}^{({\cal{C}})}_{\alpha\beta}(x,\xi,Z_1,Z_2)\phi_s,\e where ${\cal
E}^{({\cal{C}})}$ is a generalized symmetric polarization tensor
which ca be explicitly written as

\b {\cal E}^{({\cal{C}})}_{\alpha\beta}= {\cal S}\left[A
\bar{Z}_{1\alpha}\bar{Z}_{2\beta}
 +B \bar{Z}_{1\alpha}\bar{\xi}_\beta+C \bar{Z}_{2\alpha}\bar{\xi}_\beta+D \bar{\xi}_{\alpha}\bar{\xi}_\beta
 +E\theta_{\alpha\beta}\right],\e with
$$A=-\frac{\sigma}{2}+\frac{\sigma {\cal{C}}}{12({\cal{C}}-1)}\left[2(\sigma+3)\;\frac{2+{\cal{C}}}{9{\cal{C}}}+\frac{2-5{\cal{C}}}{{\cal{C}}}-(2\sigma+3)\right],$$
$$B=-\frac{1}{2}\sigma(\sigma+2)+\frac{\sigma {\cal{C}}}{12({\cal{C}}-1)}\left[2\sigma(\sigma+3)\;\frac{2+{\cal{C}}}{9{\cal{C}}}+(\sigma+2)\;\frac{2-5{\cal{C}}}{{\cal{C}}}-\sigma(2\sigma+5)\right]\frac{x.Z_2}{x.\xi},$$
$$C=\frac{\sigma {\cal{C}}}{6({\cal{C}}-1)}\left[\sigma(\sigma+3)\;\frac{2+{\cal{C}}}{9{\cal{C}}}+(\sigma+1)\;\frac{2-5{\cal{C}}}{{\cal{C}}}-\sigma(\sigma+1)\right]\frac{x.Z_1}{x.\xi},$$
$$D=\frac{\sigma(\sigma-1){\cal{C}}}{12({\cal{C}}-1)}\left[H^{-2}\left(\sigma\;\frac{2+{\cal{C}}}{9{\cal{C}}}-2-\sigma\right)\frac{Z_1.Z_2}{(x.\xi)^2}
+\left(\sigma(\sigma+3)\;\frac{2+{\cal{C}}}{9{\cal{C}}}\right.\right.
$$$$\left.\left.+(\sigma+2)\;\frac{2-5c}{c}-\sigma(\sigma+2)\right)\frac{(x.Z_1)(x.Z_2)}{(x.\xi)^2}\right],
$$
$$E=\frac{\sigma}{6}\left[\left(1+\frac{{\cal{C}}}{2({\cal{C}}-1)}\left[\sigma\;\frac{2+{\cal{C}}}{9{\cal{C}}}-\sigma-2\right]\right)Z_1.Z_2\right.
$$$$\left.+(\sigma+3)\left(1+\frac{1}{9({\cal{C}}-1)}\left(11-26{\cal{C}}+\sigma-4{\cal{C}}\sigma\right)H^2(x.Z_1)(x.Z_2)\right)\right].$$

As mentioned before, the ansatz ($3.14$) is a solution to the field
equations (3.11) with respect to $x'$ too. If we reexamine ($3.14$)
in Eq.(3.11) and treat  $x'$ as the variable instead of $x$, the
resulting equations are exactly the same as Eqs.(3.20) and (3.27)
with replacing $x$ by $x'$ and operators act on the variable $x'$
\cite{derota}. But the final expression for the tensor field
$\K^{({\cal{C}})}_{\alpha \beta}(x)$ is not other than that given in
Eq.(3.32).

We now return to the gauge-fixed value ${\cal{C}}=1$ in Eq.$(3.11)$
and reexamine the solution given in Eq.$(3.14)$, we find that
Eqs.(3.17) and (3.18) remain unchanged and Eq.(3.19) modifies as \b
Q^{(1)}_1 \left[x(Z_1.K)-(Z_1.D_1)
K-(x.Z_1)K+\frac{1}{3}D_1(Z_1.K)\right]+\frac{4}{3}D_1(Z_1.K)=0.\e
Noting Eqs.(3.24)-(3.27), one can easily confirm the validity of
Eq.$(3.33)$. It means that $K_g^{({\cal{C}})}$ can be regarded as an
arbitrary vector field without any constraint.

\subsection{The graviton two-point function}

The graviton two-point function ${\cal{ W}}_{\alpha\beta
\alpha'\beta'}(x,x')$, which is a solution of the wave equation with
respect to $x$ or $x'$, can be found in terms of the scalar
two-point function. let us try the following possibility \cite{de22,
derota}, \cite{gagata}-\cite{tap} \b {\cal W}(x,x')=\theta
\theta'{\cal W}_0(x,x')+{\cal S}{\cal S}'\theta.\theta'{\cal
W}_{1}(x,x')+D_2D'_2{\cal W}_g(x,x'),\e where ${\cal W}$, ${\cal
W}_{1}$ and ${\cal W}_{g}$ are transverse bi-vectors, ${\cal W}_{0}$
is bi-scalar and $D_2D'_2=D'_2D_2$. We now substitute the two-point
function (3.34) in the field equations (3.11) as a solution with
respect to $x$. It is easy to show that \b
Q^{(1)}_1\left(Q^{(1)}_1+2\right){\cal
W}_{1}=0,\;\;\;\;\mbox{or}\;\;\;\;Q^{(1)}_1Q^{(1)}_0{\cal
W}_{1}=0,\;\;\;\partial .{\cal W}_{1}=0=x.{\cal W}_{1},\e \b
(Q^{(1)}_0+4)(Q^{(1)}_0+6)\theta'{\cal W}_0+8(Q^{(1)}_0+2){\cal
S}'\theta'.{\cal W}_{1}=0,\e
$$(Q^{(1)}_1+4)(Q^{(1)}_1+6)D'_2{\cal W}_g^{({\cal{C}})} =\frac{4H^2}{1-{\cal{C}}}{\cal S}'\left[(Q^{(1)}_1+5)(x.\theta'){\cal W}_{1}+\theta'.D_1{\cal W}_{1}+x\theta'.{\cal W}_{1}\right]\;\;\;\;\;\;\;\;\;\;\;\;\;\;\;\;\;\;\;\;\;\;\;\; $$
\b+\frac{{\cal{C}}H^2}{1-{\cal{C}}}(Q^{(1)}_1+4)\left[D_1\theta'{\cal
W}_0+{\cal{S'}}\left(D_1(\theta'.{\cal W}_{1})-(\theta'.D_1){\cal
W}_{1}+x(\theta'.{\cal W}_{1})-5(x.\theta'){\cal W}_{1}\right)
\right], \;\;\;\;\; {\cal{C}}\neq 1. \e The solution to Eqs.(3.35)
and (3.36) are as follows  \b \theta'{\cal
W}_0(x,x')=-\frac{2}{3}{\cal S}'\theta'.{\cal W}_{1}(x,x') ,\e  \b
{\cal
W}_{1}=\left[\theta.\theta'+\frac{1}{2}D_1\left(H^2x.\theta'Q^{(1)}_0-\theta'.\bar{\partial}-2H^2x.\theta'
\right)\right]{\cal W}_{s},\e where ${\cal W}_{s}$ is the two-point
function for dS massless minimally coupled scalar field, obtained
from ``Gupta-Bleuler vacuum'' state, with the following explicit
form \cite{ta3} \b {\cal W}_{s}(x,x')=\frac{iH^2}{8\pi^2} \epsilon
(x^0-x'^0)[\delta(1-{\cal Z}(x,x'))+\vartheta ({\cal Z}(x,x')-1)],
\e with \b {\cal{Z}}=-H^2x.x',\;\;\;\;\;\; \mbox{and} \;\;\;\;\;\; \epsilon (x^0-x'^0)=\left\{ \ba{rcl} 1&x^0>x'^0 ,\\
0&x^0=x'^0 ,\\ -1&x^0<x'^0.\\ \ea\right.\e which preserves
dS-invariance.

Combining Eq.s $(3.37)$ and (3.38) we have
$$D'_2{\cal W}_g^{({\cal{C}})}=\frac{H^2}{3(1-{\cal{C}})} {\cal S}'\left[(x'.\theta){\cal
W}_{1}+\frac{1}{9}D'_1(\theta.{\cal W}_{1})\right]$$
\b+\frac{{\cal{C}}H^2}{1-{\cal{C}}}(Q^{(1)}_1+6)^{-1}{\cal
S}'\left[\frac{1}{3}D_1(\theta'.{\cal W}_{1})-(\theta'.D_1){\cal
W}_{1}+x(\theta'.{\cal W}_{1})-5(x.\theta'){\cal W}_{1} \right],\e
from which we obtain \b D'_2{\cal
W}^{({\cal{C}})}_g(x,x')=\frac{{\cal{C}}H^2}{6(1-{\cal{C}})} {\cal
S}' \left[\frac{2+{\cal{C}}}{9{\cal{C}}}D_1\theta'.{\cal W}_{1}
+\frac{2-5{\cal{C}}}{{\cal{C}}}x.\theta'{\cal W}_{1}+x\theta'.{\cal
W}_{1} -H^{-2}\theta'.\bar \partial{\cal W}_1\right].\e Making use
of Eqs.$(3.38)$, $(3.39)$ and $(3.43)$, after some relatively simple
and straightforward calculations, it turns out that the tensor
two-point function can be written in the form
$$ {\cal W}^{({\cal{C}})}_{\alpha\beta \alpha'\beta'}(x,x')=\frac{2{\cal{Z}}}{27(1-{\cal{C}})(1-{\cal{Z}}^2)^2}{\cal
S}{\cal S}'
\left[P_1({\cal{C}},{\cal{Z}})\theta_{\alpha\beta}\theta'_{\alpha'\beta'}\right.
+P_2({\cal{C}},{\cal{Z}})(\theta_{\alpha}.\theta'_{\alpha'})(\theta_{\beta}.\theta'_{\beta'})$$
$$+H^2P_3({\cal{C}},{\cal{Z}})\left(\theta_{\alpha\beta}(x.\theta'_{\alpha'})(x.\theta'_{\beta'})
+\theta'_{\alpha'\beta'}(x'.\theta_{\alpha})(x'.\theta_{\beta})\right)
+P_4({\cal{C}},{\cal{Z}})H^4(x'.\theta_{\alpha})(x'.\theta_{\beta})(x.\theta'_{\alpha'})(x.\theta'_{\beta'})$$
\b\left.+P_5({\cal{C}},{\cal{Z}})H^2(\theta_{\alpha}.\theta'_{\alpha'})(x.\theta'_{\beta'})(x'.\theta_{\beta})
\right]\frac{d }{d{\cal{Z}}}{\cal W}_{s}({\cal{Z}}),\e where
$$P_1({\cal{C}},{\cal{Z}})=(1-{\cal{Z}}^2)\left[2+{\cal{C}}+3({\cal{C}}-1){\cal{Z}}^2\right],$$
$$P_2({\cal{C}},{\cal{Z}})=(1-{\cal{Z}}^2)\left[17{\cal{C}}-11+9(1-{\cal{C}}){\cal{Z}}^2\right],\;\;\;\;\;\;P_3({\cal{C}},{\cal{Z}})=3\left[7{\cal{C}}-1+({\cal{C}}-1){\cal{Z}}^2 \right],$$
$$P_4({\cal{C}},{\cal{Z}})=-\frac{3}{(1-{\cal{Z}}^2)}\left[3(7-19{\cal{C}})-2(1+5{\cal{C}}){\cal{Z}}^2-3(1-{\cal{C}}){\cal{Z}}^4\right],$$
$$P_5({\cal{C}},{\cal{Z}})=\frac{1}{{\cal{Z}}}\left[10(4-7{\cal{C}})+2(1-22{\cal{C}}){\cal{Z}}^2-18(1-{\cal{C}}){\cal{Z}}^4\right].$$
The Eq.$(3.44)$ is the explicit form of the graviton two-point
function, based on the quadratically-extended gravitational theory,
in the ambient space notations. It is clearly dS-invariant and free
of any divergences. It is well-known that if we require the
statement $(3.34)$ to be the solution of Eq.$(3.11)$, with respect
to $x'$ as the variable, the final result is not other than
$(3.44)$.

Let's translate the two-point function $(3.44)$ to the dS intrinsic
coordinates. Using the translation rules given in \cite{derota,
gagata, de} we have
$$
W^{({\cal{C}})}_{aba'b'}(X,X')=\frac{2{\cal{Z}}}{27(1-{\cal{C}})}{\cal
S}{\cal S}'\left[\frac{P_1}{(1-{\cal{Z}}^2)^2}\;g_{ab}g'_{a'b'}
+\frac{P_2}{(1-{\cal{Z}}^2)^2}\;g_{aa'}g_{bb'}\right.$$
$$+\frac{P_3}{1-{\cal{Z}}^2}\;\left(g_{ab}n_{a'}n_{b'}+g'_{a'b'}n_{a}
n_{b}\right)+\left(\frac{2({\cal{Z}}-1)P_2}{(1-{\cal{Z}}^2)^2}
+\frac{P_5}{1-{\cal{Z}}^2} \right)g_{aa'}n_bn_{b'}
$$\b\left.+\left(\frac{P_2}{(1+{\cal{Z}})^2}-\frac{P_5}{1+{\cal{Z}}}+P_4
\right)n_{a} n_{b}n_{a'}n_{b'}\right]\frac{d }{d{\cal{Z}}}{\cal
W}_{s}({\cal{Z}}).\e

If we putt ${\cal{C}}=0$ the two-point functions $(3.44)$ and
$(3.45)$ will reduce to the corresponding physical two-point
functions. It is notable that $W^{({\cal{C}})}_{aba'b'}(X,X')$ and
${\cal W}^{({\cal{C}})}_{\alpha\beta \alpha'\beta'}(x,x')$ are
related through the following tensorial transformation rule
 \b
W^{{\cal{C}}}_{aba'b'}(X,X')=\frac{\partial x^{\alpha}}{\partial
X^{a}}\frac{\partial x^{\beta}}{\partial X^{b}}\frac{\partial
x'^{\alpha'}}{\partial X'^{a'}}\frac{\partial x'^{\beta'}}{\partial
X'^{b'}}{\cal W}^{{\cal{C}}}_{\alpha\beta\alpha'\beta'}(x,x'). \e

\section{Conclusion }

This work considers an extension of the Einstein-Hilbert
gravitational action, which is constructed out by the linear
combination of Ricci scalar, Ricci invariant and Riemann invariant
in dS space. Varying the proposed action with resect to metric
tensor leads to the fourth order field equations, conventionally
named as the quadratically-extended gravitational field equations
(Eq.(2.3)). The background field method is utilized and the
linearized field equations are obtained in terms of intrinsic
coordinates in the four-dimensional dS space as the background
(Eq.(2.14)). The linearized field equations is invariant under some
gauge transformations. We obtained the Minkowskian correspondence of
the theory by taking the zero curvature limit (Eq.(2.16)). It
remains invariant under some gauge transformations too.

By imposing some simple conditions (Eq.(2.18)), the massless spin-2
field equations has been written in the ambient space notations with
the Casimir operators appearing in it (Eq.(3.7)). It is also gauge
invariant under some special gauge transformations. Because of this
gauge freedom a gauge-fixing parameter ${\cal{C}}$ is inserted in
the field equations. The field equations, in the physical state, are
correspond to two of the UIR's of dS group denoted by
$\Pi^{\pm}_{2,2}$ and $\Pi^{\pm}_{2,1}$ in discrete series. We
believe that for a model of gravity theory to be valid and
successful it is necessary for the field equations to transform
according to the UIR's of the symmetric group \cite{de22}. With this
issue in mind we introduced a new four-dimensional GB-like action
(Eq.(3.13)), which is expected to be a successful model of modified
gravity theory and have some new and realized physical consequences
in the context of quantum gravity toy models.

Next we tried to illustrate the validity and successfulness of the
gravitational model we just obtained. For this purpose we solved the
related field equations. The symmetric tensor field, as the solution
of the field equations, has been obtained in terms of the
gauge-fixing parameter ${\cal{C}}$. It has been written as the
multiplication of a symmetric rank-2 generalized polarization tensor
and a massless minimally coupled scalar field in dS space
(Eq.(3.31). Also, we have calculated the relevant two-point
function, making use of the ambient formalism. We showed that the
two-point function can be written in terms of the massless minimally
coupled scalar two-point function in dS space (Eq.(3.44)). It is
dS-invariant and free of any theoretical problems. The graviton
two-point function has been written in terms of dS intrinsic
coordinates from its ambient space counterpart which is dS-invariant
and free of any theoretical problems too (Eq.(3.45)). Therefore, we
claim that the proposed model of modified gravitational theory (i.e.
the four-dimensional GB-like action given in Eq.(3.13)) is a valid
and successful model and can produce reasonable and interesting
results in the contexts of classical theory of gravity and quantum
gravity toy models. It must be stressed that the results of this
work confirm that it is necessary for a theory of modified gravity
to be successful if it transforms according to the UIR's of the
related symmetric group. It is compatible with the results of
\cite{de22}.

\setcounter{equation}{0}
\begin{appendix}
\section{Some useful mathematical relations }
The following relations have been used in deriving the linearized
field equations.\b\label {eq:LG 28}
\tilde{R}_{abcd}=H^{2}(\tilde{g}_{ac}\tilde{g}_{bd}-\tilde{g}_{ad}\tilde{g}_{bc}),\e
\b\label {eq:LG 39} \tilde{R}_{ab}=3H^{2}\tilde{g}_{ab},\e \b \label
{eq:LG 40}\tilde{R}=12H^{2},\e
  \b\label {eq:LG
34}({\cal{R}})_L=\nabla_{c}\nabla_{b}h^{cb}-\Box h'-3H^2h'.\e \b
(\nabla_a \nabla_b {\cal{R}})_L = \nabla_a \nabla_b \left(
\nabla_c\nabla_d h^{cd}-\Box h'-3H^2 h'\right).\e \b (\Box
{\cal{R}})_L=\Box \left( \nabla_c\nabla_d h^{cd}-\Box h'-3H^2
h'\right).\e
 \b
({\cal{R}}^c_{d})_L =\frac{1}{2}\left(\nabla^{c}\nabla_{a}h_{d}^{a}
+\nabla_d\nabla_{a}h^{ac}+8H^2h^c_{d}-2H^2h'\tilde{g}^c_{d}-\Box
h^c_{d}-\nabla^{c}\nabla_{d}h'\right)-3H^2h^{c}_{d}.\e  \b (
{\cal{R}}^{bc})_L=\frac{1}{2}\left(\nabla^{c}\nabla_{a}h^{ab}
+\nabla^b\nabla_{a}h^{ac}+8H^2h^{bc}-2H^2h'\tilde{g}^{bc}-\Box
h^{bc}-\nabla^{c}\nabla^{b}h'\right)-6H^2h^{bc}.\e \b\label {eq:LG
32}
({\cal{R}}_{ab})_L=\frac{1}{2}\left(\nabla_{a}\nabla_{c}h_{b}^{c}
+\nabla_{b}\nabla_{c}h_{a}^{c}+8H^2h_{ab}-2H^2h'\tilde{g}_{ab}-\Box
h_{ab}-\nabla_{a}\nabla_{b}h'\right).\e
 \b (\nabla_a \nabla_b
{\cal{R}}_{cd})_L= \frac{1}{2} \nabla_a \nabla_b
\left(\nabla_{c}\nabla_{e}h_{d}^{e}
+\nabla_{d}\nabla_{e}h_{c}^{e}+2H^2h_{cd}-2H^2h'\tilde{g}_{cd}-\Box
h_{cd}-\nabla_{c}\nabla_{d}h'\right) .\e \b (\nabla_a \nabla_b
{\cal{R}}^c_{d})_L= \frac{1}{2} \nabla_a \nabla_b
\left(\nabla^{c}\nabla_{e}h_{d}^{e}
+\nabla_{d}\nabla_{e}h^{ce}+2H^2h^c_{d}-2H^2h'\tilde{g}^c_{d}-\Box
h^c_{d}-\nabla^{c}\nabla_{d}h'\right),\e \b (\nabla_a \nabla_b
{\cal{R}}^{cd})_L= \frac{1}{2} \nabla_a \nabla_b
\left(\nabla^{c}\nabla_{e}h^{ed}
+\nabla^{d}\nabla_{e}h^{ce}+2H^2h^{cd}-2H^2h'\tilde{g}^{cd} -\Box
h^{cd}-\nabla^{c}\nabla^{d}h'\right).\e \b (\Box
{\cal{R}}_{cd})_L=\frac{1}{2} \Box
\left(\nabla_{c}\nabla_{e}h_{d}^{e}
+\nabla_{d}\nabla_{e}h_{c}^{e}+2H^2h_{cd}-2H^2h'\tilde{g}_{cd}-\Box
h_{cd}-\nabla_{c}\nabla_{d}h'\right).\e
 \b ({\cal{R}}^{c}\;_{dab})_L=\frac{1}{2}\left[\nabla_{a}\left(\nabla_{d}h_{b}^{c}
+\nabla_{b}h_{d}^{c}-\nabla^{c}h_{db}\right)-\nabla_{b}\left(\nabla_{d}h_{a}^{c}
+\nabla_{a}h_{d}^{c}-\nabla^{c}h_{ad} \right) \right].\e
 \b ({\cal{R}}_{acde})_L=\frac{1}{2}\left[\nabla_{d}\left(\nabla_{c}h_{ae}+\nabla_{e}h_{ac}-\nabla_{a}h_{ce}
\right)-\nabla_{e}\left(\nabla_{c}h_{ad}+\nabla_{d}h_{ca}-\nabla_{a}h_{cd}
\right) \right]+g_{ec}h_{ad}-g_{cd}h_{ae}.\e \b
({\cal{R}}_b^{\;\;cde})_L=\frac{1}{2}\left[\nabla^{d}\left(\nabla^{c}h_{b}^e+\nabla^{e}h_{b}^c-\nabla_{b}h^{ce}
\right)-\nabla^{e}\left(\nabla^{c}h_{b}^d+\nabla^{d}h_{b}^c-\nabla_{b}h^{cd}
\right) \right]+2\left(g_b^{e}h^{cd}-g_b^{d}h^{ce}\right).\e
$$ ({\cal{R}}^{cdef})_L=\frac{1}{2}\left[\nabla^{e}\left(\nabla^{d}h^{cf}+\nabla^{f}h^{cd}-\nabla^{c}h^{fd}
\right)-\nabla^{f}\left(\nabla^{d}h^{ec}+\nabla^{e}h^{cd}-\nabla^{c}h^{ed}
\right) \right]$$ \b+2\left(
g^{cf}h^{ed}-g^{ce}h^{fd}\right)+g^{ed}h^{cf}-g^{fd}h^{ce}.\e

\setcounter{equation}{0}
\section{Details of derivations of Eq.(3.7)}

In order to write the field equations in terms of the five
dimensional ambient space notations, we need the following
identities \b D_\alpha D_\gamma
\K^\gamma_\beta=\bar{\partial}_\alpha(\bar{\partial}.\K)_\beta-H^2\K'\eta_{\alpha\beta}
-H^2x_{\beta}\bar{\partial}_\alpha\K'-H^2x_{\beta}(\bar{\partial}.\K)_\alpha
,\e \b \Box_A \K_{\alpha \beta}= D_\gamma D^\gamma \K_{\alpha
\beta}=-H^2Q^{(1)}_0\K_{\alpha \beta}-2H^2\K_{\alpha
\beta}-2H^2{\cal{S}}x_\alpha(\bar{\partial}.\K)_\beta+2H^4x_\alpha
x_\beta \K',\e \b D_\alpha D_\beta ( \Box_A \K')= D_\alpha D_\beta (
D_\gamma D^\gamma \K')=-H^2\bar{\partial}_\alpha\bar{\partial}_\beta
Q^{(1)}_0\K' +H^4x_\beta \bar{\partial}_\alpha Q^{(1)}_0\K',\e \b
D_\alpha D_\beta ( D_\gamma D_\lambda
\K^{\gamma\lambda})=\bar{\partial}_\alpha \bar{\partial}_\beta
\bar{\partial}.(\bar{\partial}.\K)-H^2x_\beta \bar{\partial}_\alpha
\bar{\partial}.(\bar{\partial}.\K)-4H^2 \bar{\partial}_\alpha
\bar{\partial}_\beta \K'+4H^4 x_\beta \bar{\partial}_\alpha \K',\e
\b D_\alpha D_\beta
\K'=\bar{\partial}_\alpha\bar{\partial}_\beta\K'-H^2x_\beta
\bar{\partial}_\alpha \K' ,\e \b \Box_A \K'= D_\alpha D^\alpha
\K'=-H^2Q^{(1)}_0\K',\e \b \Box^2_A \K'= (D_\alpha D^\alpha
)^2\K'=H^4(Q^{(1)}_0)^2\K',\e \b D_\alpha D_\beta
\K^{\alpha\beta}=\bar{\partial}.(\bar{\partial}.\K)-4H^2\K',\e
$$ H^{-4}\Box^2_A \K_{\alpha \beta}=H^{-4} (D_\gamma D^\gamma
)^2\K_{\alpha \beta}= (Q^{(1)}_0)^2 \K_{\alpha \beta}+4Q^{(1)}_0
\K_{\alpha \beta}+4 \K_{\alpha \beta}$$
$$ +{\cal{S}}\left[4x_\alpha Q^{(1)}_0(\bar{\partial}.\K)_\beta+8x_\alpha (\bar{\partial}.\K)_\beta
-14H^2x_\alpha x_\beta
\K'-4H^{-2}\bar{\partial}_\alpha(\bar{\partial}.\K)_\beta \right.$$
\b \left.+4 x_\alpha x_\beta
\bar{\partial}.(\bar{\partial}.\K)-2H^{-2}x_\alpha x_\beta Q^{(1)}_0
\K'+4 x_\alpha \bar{\partial}_\beta \K'+2\theta_{\alpha\beta}\K'
\right],\e
$$\Box_A {\cal{S}}\left(D_\alpha D_\gamma
\K^\gamma_\beta\right)=-H^2{\cal{S}}\left[\bar{\partial}_\alpha
Q^{(1)}_0(\bar{\partial}.\K)_\beta+6\bar{\partial}_\alpha
(\bar{\partial}.\K)_\beta-H^2x_\alpha Q^{(1)}_0
(\bar{\partial}.\K)_\beta+2x_\alpha \bar{\partial}_\beta
\bar{\partial}.(\bar{\partial}.\K)\right.$$ $$-2H^2x_\alpha x_\beta
\bar{\partial}.(\bar{\partial}.\K)-12 H^2x_\alpha
\bar{\partial}_\beta
\K'-6H^2x_\alpha(\bar{\partial}.\K)_\beta+10H^4x_\alpha x_\beta
\K'$$ \b \left. -H^2\eta_{\alpha\beta}Q^{(1)}_0
\K'-2H^2\theta_{\alpha\beta} \K'-H^2x_\alpha \bar{\partial}_\beta
Q^{(1)}_0
\K'+2\bar{\partial}_\alpha\bar{\partial}_\beta\K'\right].\e In
obtaining the above identities, Eq.(3.2) has been used.

Now substituting Eqs.(B.1)-(B.10) in Eq.(3.10) and adopting the
metric signature $(+, -, -, -)$ results in
$$ Q^{(1)}_0(Q^{(1)}_0-2)\K_{\alpha\beta}+{\cal{S}}\left[3x_\alpha Q^{(1)}_0(\bar{\partial}.\K)_\beta
+H^{-2}\bar{\partial}_\alpha
Q^{(1)}_0(\bar{\partial}.\K)_\beta-6x_\alpha
(\bar{\partial}.\K)_\beta +2H^2x_\alpha x_\beta \K' \right. $$
$$ -2H^{-2}\bar{\partial}_\alpha(\bar{\partial}.\K)_\beta +2x_\alpha x_\beta\bar{\partial}.(\bar{\partial}.\K)
-2H^2x_\alpha x_\beta  Q^{(1)}_0\K'-4x_\alpha \bar{\partial}_\beta
\K'+2H^{-2}x_\alpha \bar{\partial}_\beta
\bar{\partial}.(\bar{\partial}.\K)$$
$$\left.-\eta_{\alpha\beta}Q^{(1)}_0\K'+2H^{-2}\bar{\partial}_\alpha\bar{\partial}_\beta \K'
-x_\beta \bar{\partial}_\alpha Q^{(1)}_0\K'+4
\eta_{\alpha\beta}\K'\right]$$ $$
-\frac{1}{3}\left(H^{-2}\bar{\partial}_\alpha\bar{\partial}_\beta-x_\beta\bar{\partial}_\alpha
\right)\left[14\K'+Q^{(1)}_0\K'-2H^{-2}\bar{\partial}.(\bar{\partial}.\K)
\right]$$
\b+\theta_{\alpha\beta}\left[2H^{-2}\bar{\partial}.(\bar{\partial}.\K)-10\K'
+\frac{7}{3}Q^{(1)}_0\K'-\frac{1}{3}H^{-2}Q^{(1)}_0\bar{\partial}.(\bar{\partial}.\K)-\frac{1}{3}\left(Q^{(1)}_0\right)^2\K'
\right].\e On the other hand, making use of Eq.(3.6) we can show
that
$$(Q_2^{(1)}+4)(Q_2^{(1)}+6)\K_{\alpha\beta}=Q_0^{(1)}\left(Q_0^{(1)}-2\right)\K_{\alpha\beta}+4Q_0^{(1)}\K'\eta_{\alpha\beta}+8\K'\eta_{\alpha\beta}$$
\b+4{\cal{S}}\left[-H^2x_\alpha x_\beta\K'+H^2x_\alpha x_\beta
\bar{\partial}.(\bar{\partial}.\K)+x_\alpha
Q_0^{(1)}(\bar{\partial}.\K)_\beta +x_\alpha \bar{\partial}_\beta
\K'-x_\alpha (\bar{\partial}.\K)_\beta-H^{-2}\bar{\partial}_\alpha
(\bar{\partial}.\K)_\beta\right].\e
 Also one can show that
\b\left(D_2(\partial_2.\K)\right)_{\alpha\beta}=
H^{-2}{\cal{S}}\left[-H^2 x_\alpha \bar{\partial}_\beta \K'-H^2
x_\alpha (\bar{\partial}.\K)_\beta+\bar{\partial}_\alpha
(\bar{\partial}.\K)_\beta \right]
-2\K'\eta_{\alpha\beta}-H^{-2}\bar{\partial}_\alpha\bar{\partial}_\beta\K'+x_\beta\bar{\partial}_\alpha\K',\e
and
$$(Q_2^{(1)}+4)\left(D_2(\partial_2.\K)\right)_{\alpha\beta}={\cal{S}}\left[-x_\alpha
Q_0^{(1)}(\bar{\partial}.\K)_\beta+H^{-2}\bar{\partial}_\alpha
Q^{(1)}_0(\bar{\partial}.\K)_\beta-2x_\alpha
(\bar{\partial}.\K)_\beta+6H^2x_\alpha x_\beta \K' \right.$$
$$ +2H^{-2}\bar{\partial}_\alpha(\bar{\partial}.\K)_\beta -2x_\alpha x_\beta\bar{\partial}.(\bar{\partial}.\K)
-2H^2x_\alpha x_\beta  Q^{(1)}_0\K'-6x_\alpha \bar{\partial}_\beta
\K'+2H^{-2}x_\alpha \bar{\partial}_\beta
\bar{\partial}.(\bar{\partial}.\K)$$ \b
\left.-3\eta_{\alpha\beta}Q^{(1)}_0\K'-x_\alpha \bar{\partial}_\beta
Q^{(1)}_0\K'\right]-\theta_{\alpha\beta}\left(12\K'-4Q^{(1)}_0\K'\right)
+\left(-H^{-2}\bar{\partial}_\alpha\bar{\partial}_\beta+x_\beta\bar{\partial}_\alpha
\right)Q^{(1)}_0\K'.\e Regarding Eqs.(B.12) and (B.14), the field
equations (B.11) reduces to
$$(Q_2^{(1)}+4)\left[(Q_2^{(1)}+6)\K_{\alpha\beta}+\left(D_2(\partial_2.\K)\right)_{\alpha\beta}\right] $$
$$-\frac{1}{3}\left(-H^{-2}\bar{\partial}_\alpha\bar{\partial}_\beta+x_\beta\bar{\partial}_\alpha
\right)\left[2Q^{(1)}_0\K'+2H^{-2}\bar{\partial}.(\bar{\partial}.\K)-2\K'
\right]$$\b+\theta_{\alpha\beta}\left[2H^{-2}\bar{\partial}.(\bar{\partial}.\K)+2\K'
-\frac{5}{3}Q^{(1)}_0\K'-\frac{1}{3}H^{-2}Q^{(1)}_0\bar{\partial}.(\bar{\partial}.\K)-\frac{1}{3}\left(Q^{(1)}_0\right)^2\K'
\right]=0.\e

Now, Eq.(B.15) can be rewritten as \b
T_{\alpha\beta}+{\cal{T}}_{\alpha\beta}-\frac{1}{4}\theta_{\alpha\beta}\left(T'+{\cal{T}}'\right)=0,\e
where \b
T_{\alpha\beta}=(Q_2^{(1)}+4)\left[(Q_2^{(1)}+6)\K_{\alpha\beta}+\left(D_2(\partial_2.\K)\right)_{\alpha\beta}\right],\e
\b
T'=\mbox{trace}\left[T_{\alpha\beta}\right]=2\left(Q^{(1)}_0\right)^2\K'+6Q^{(1)}_0K'-8\K'-8H^{-2}\bar{\partial}.(\bar{\partial}.\K)
+2H^{-2}Q^{(1)}_0\bar{\partial}.(\bar{\partial}.\K),\e \b
{\cal{T}}_{\alpha\beta}=D_{\alpha\beta}{\cal{T}},\;\;D_{\alpha\beta}=-1/3\left(-H^{-2}\bar{\partial}_\alpha\bar{\partial}_\beta
+x_\beta\bar{\partial}_\alpha \right) \mbox{is a symmetric tensor
operator},\e
\b{\cal{T}}=\mbox{trace}\left[(Q_2^{(1)}+6)\K_{\alpha\beta}+\left(D_2(\partial_2.\K)\right)_{\alpha\beta}\right]
=2Q^{(1)}_0\K'-2K'+2H^{-2}\bar{\partial}.(\bar{\partial}.\K),\e \b
{\cal{T}}'=\mbox{trace}\left[{\cal{T}}_{\alpha\beta}\right]=-\frac{2}{3}\left[\left(Q^{(1)}_0\right)^2\K'-Q^{(1)}_0K'
+H^{-2}Q^{(1)}_0\bar{\partial}.(\bar{\partial}.\K)\right].\e Now,
the field equations (B.16) can be written as  \b
{\cal{O}}\left[(Q_2^{(1)}+4)(Q_2^{(1)}+6)\K_{\alpha\beta}+(Q_2^{(1)}+4)\left(D_2(\partial_2.\K)\right)_{\alpha\beta}\right]=0
,\e where ${\cal{O}}$ is an operator with the operation
\b{\cal{O}}T_{\alpha\beta}=T_{\alpha\beta}+{\cal{T}}_{\alpha\beta}-\frac{1}{4}\theta_{\alpha\beta}(T'+{\cal{T}}').\e
Since ${\cal{O}}$ has an inverse Eq.(B.22) is nothing but Eq.(3.7).

\end{appendix}

\end{document}